\documentclass[twocolumn,aps,prc,tightenlines,floats,floatfix,nofootinbib]{revtex4}

\def\bea{\begin{eqnarray}}
\def\eea{\end{eqnarray}}

\def\sfrac#1#2{{\textstyle \frac{#1}{#2}}}

\def\be{\begin{equation}}
\def\ee{\end{equation}}
\def\ba{\begin{eqnarray}}
\def\ea{\end{eqnarray}}

\def\mb{\mbox}
\def\bm{\boldmath}

\usepackage{graphics}
\usepackage{graphicx}
\usepackage{epsf}
\usepackage{amsmath}
\usepackage{amssymb}

\begin{document}

\phantom{0}
\vspace{-0.2in}
\hspace{5.5in}
\parbox{1.5in}{ \leftline{JLAB-THY-10-1219}}

\vspace{-1in}

\title
{\bf A model for the $\Delta(1600)$ resonance and
$\gamma N \to \Delta(1600)$ transition}

\author{G.~Ramalho$^1$ and K.~Tsushima$^2$ \vspace{-0.15in}}

\affiliation{
$^1$Centro de F{\'\i}sica Te\'orica de Part{\'\i}culas,
Av.\ Rovisco Pais, 1049-001 Lisboa, Portugal \vspace{-0.15in}}

\affiliation{
$^2$Excited Baryon Analysis Center (EBAC) in Theory Center,
Thomas Jefferson National Accelerator Facility, Newport News, VA 23606, USA
}

\vspace{0.2in}
\date{\today}

\phantom{0}

\begin{abstract}
A covariant spectator constituent quark model is applied to study the
$\gamma N \to \Delta(1600)$ transition.
Two processes are important in the transition:
a photon couples to the individual quarks of the $\Delta(1600)$
core (quark core),
and a photon couples to the intermediate pion-baryon states (pion cloud).
While the quark core contributions are estimated
assuming $\Delta(1600)$ as the first radial excitation of $\Delta(1232)$,
the pion cloud contributions are estimated based on an analogy with
the $\gamma N \to \Delta(1232)$ transition.
To estimate the pion cloud contributions in the
$\gamma N \to \Delta(1600)$ transition,
we include the relevant intermediate states,
$\pi N$, $\pi \Delta$, $\pi N(1440)$ and $\pi \Delta(1600)$.
Dependence on the four-momentum transfer squared, $Q^2$,
is predicted for the magnetic dipole transition form factor,
$G_M^\ast(Q^2)$, as well as the helicity amplitudes,
$A_{1/2}(Q^2)$ and $A_{3/2}(Q^2)$. The results at $Q^2=0$ are compared
with the existing data.
\end{abstract}

\vspace*{0.9in}  
\maketitle

\section{Introduction}

With the extended energy range and the increased
precision of recent accelerators,
it is now possible to probe the electromagnetic structure
of baryon resonances beyond the first resonance region.
For example, the facilities like CLAS (Jefferson Lab),
MAMI (Mainz), ELSA (Boon),
LEGS (Brookhaven), BATES (MIT) and Spring-8 (Japan),
are able to measure the electromagnetic
transition properties associated with the
resonances such as $P_{33}(1232)$, $P_{11}(1440)$, $S_{11}(1535)$
and $D_{13}(1520)$~\cite{Burkert04}.

To accommodate the data with higher four-momentum transfer squared $Q^2$
and center-of-mass energy $W$, an improvement of the data analysis is mandatory,
since more baryons and mesons emerge in the possible intermediate states.
Then, the exact identification of the resonances
easily becomes controversial due to a large number of channels
taking place in a very small energy range.
Furthermore, the analysis made by various groups,
MAID~\cite{Drechsel07},
SAID~\cite{Arndt08},
JLab-Yeveran~\cite{Aznauryan07,CLAS}
JLMS (JLab-Moscow)~\cite{Mokeev09},
Bonn-Gatchina \cite{Bonn},
Valencia~\cite{Doring06},
Giessen~\cite{Penner02}, CMB~\cite{CMB}
and KSU~\cite{KSU}, are based on
different resonance poles and meson-baryon coupling constants.
A similar situation exists also for
the dynamical coupled-channel reaction models, e.g.,
Mainz-Taipei~\cite{Kamalov01}, J\"{u}lich~\cite{Doring09a,Julich}
and EBAC~\cite{SatoLee,Diaz07c,Matsuyama07,Diaz07a} models.
One of the most known issues which drives them
to study the baryon resonances is "missing resonance problem", that is,
several states predicted by quark models have not yet been observed nor
identified by experiments~\cite{Burkert04,Capstick92a}.

Under such circumstances
any theoretical studies for the baryon resonances
would help to uncover their properties.
For example, $\Delta(1600)$,
the first excited state of $\Delta$,
may be one of such interesting resonances worth while to
study. It has not yet been studied well so far.
According to the Particle Data Group (PDG)~\cite{PDG}
$\Delta(1600)$ is a three-star resonance
which decays preferentially to $\pi N$, $\pi N(1440)$ and
$\pi \Delta$.
Although the form factors associated with
the electroproduction of this resonance were studied
some time ago by a constituent quark model~\cite{Capstick95},
it is only recent that this resonance has been
included in meson-baryon reaction
analysis models~\cite{Burkert04,Bonn,Golli08b,CLAS}.
Recent experiments show that the $\Delta(1600)$ resonance
can be very important in double pion production
in nucleon-nucleon collisions~\cite{Skorodko09b,Skorodko10a,Cao10a}.
The $\Delta(1600)$ resonance was also studied in
lattice QCD simulations~\cite{Leinweber04,Engel09}
and QCD sum rules~\cite{Erkol08}.

Although the experimental access for the $\Delta(1600)$ resonance
is still insufficient, it is very interesting to study
it theoretically by the following reasons.
Similarly to its ground state $\Delta(1232)$, $\Delta(1600)$
can be described as a quark core
dressed by meson cloud in the low $Q^2$ region.
However, contrarily to the $\Delta(1232)$ case,
meson cloud structure for the $\Delta(1600)$
is expected to be much richer, since more meson-baryon channels
are associated with it~\cite{PDG}.
To estimate the meson-baryon dressing for the $\Delta(1600)$,
one can in principle use a dynamical coupled-channel models,
but it is also necessary to understand
the three-quark core structure based on the
underlying physics of QCD, instead of using
a phenomenological parametrization.
Thus, the use of a quark model,
which includes the degrees of freedom that
dominate in the intermediate and higher $Q^2$ region generally,
is a natural consequence.
Quark models have also proven to be
very useful in the studies of nucleon and $\Delta(1232)$ systems.

In this article we study
the structure of $\Delta(1600)$ and
the $\gamma N \to \Delta(1600)$ transition,
by applying a covariant spectator constituent
quark model~\cite{Nucleon,FixedAxis,NDelta,NDeltaD,LatticeD,Roper,Lattice,DeltaFF0,DeltaFF,Omega,Octet}.
In this model one can naturally assume that the $\Delta(1600)$
as the first radial excitation of $\Delta(1232)$,
similarly to the case of $N(1440)$ and $N$~\cite{Roper}.
The interpretation of the $\Delta(1600)$ resonance as the
first radial excitation of $\Delta(1232)$
is sufficient for the present approach to
estimate the valence quark contributions for
the $\gamma N \to \Delta(1600)$ transition.
No extra parameters are required.
However, the  transitions $\gamma N \to \Delta(1232)$
and $\gamma N \to \Delta(1600)$, are
very different with respect to the pion cloud effects.
While the pion cloud contributions are about 30-45\%
for the transition with the $\Delta(1232)$,
the pion cloud contributions would be more
significant for the transition with the $\Delta(1600)$.
This is due to the increase of relevant intermediate states,
$\pi N$, $\pi \Delta $, $\pi N(1440)$ and $\pi \Delta(1600)$.

Before discussing any details of the numerical results,
we can state that, solely from the quark
core contributions, the magnetic dipole transition form factors
in the $\gamma N \to \Delta(1600)$ transition at $Q^2=0$
is negative ($-1.11$), and significantly undershoot
the experimental data ($\approx +0.20$).
However, with the inclusion of the pion cloud
contributions which are dominant in the low $Q^2$ region,
the final result approaches to the experimental data points.

This article is organized as follows.
In Sec.~\ref{secReaction} general remarks are given for the
$\gamma N \to \Delta$ transitions,
and the theoretical background is introduced.
In Secs.~\ref{secValence} and~\ref{secPion}
contributions from the valence
quarks and pion cloud are respectively discussed
for the $\gamma N \to \Delta(1600)$ transition.
In Sec.~\ref{secHelAmp} reaction observables are
discussed, and their results are presented in Sec.~\ref{secResults}.
The conclusion is given in Sec.~\ref{secConclusions}.

\section{General remarks on the $\gamma N \to \Delta$ transition}
\label{secReaction}

The electromagnetic transition between a
spin 1/2 baryon (e.g., nucleon) and a spin 3/2 baryon (e.g., $\Delta$)
with the positive parity,
can be described in terms of three independent
form factors introduced by Jones and Scadron~\cite{Jones73}:
$G_M^\ast$ (magnetic dipole),
$G_E^\ast$ (electric quadrupole)
and $G_C^\ast$ (Coulomb quadrupole).
An example is the $\gamma N \to \Delta(1232)$ transition.

It is well established
that the $\gamma N \to \Delta(1232)$ transition is dominated by
the magnetic dipole form factor,
$G_M^\ast$~\cite{Burkert04,NDelta,NDeltaD,Becchi65,Isgur82,Pascalutsa07,Villano09}.
This can be easily understood
in a naive SU(6) quark model, since $\Delta(1232)$ can be regarded as
a system which one quark spin is flipped from the nucleon system,
therefore possible by a pure magnetic dipole transition.
In fact, the magnetic dipole form factor, $G_M^\ast$,
emerges naturally as a dominant form factor when
only S-state structure in the quark-diquark system is included for
the $\Delta(1232)$ wave function,
and all the remaining form factors vanish~\cite{NDelta,Becchi65,Isgur82}.
Although D-states can contribute to the transition, they
induce only small corrections for $G_M^\ast$,
besides the contribution for the quadrupole form factors,
which are also small when compared with $G_M^\ast$~\cite{NDeltaD,LatticeD}.

However, it is well known that only the valence quark
contributions are insufficient to describe the
$\gamma N \to \Delta(1232)$
transition~\cite{Burkert04,NDelta,Pascalutsa07,Faessler06,Rohrwild07,Braun06,Wang09}.
The pion cloud contributions, which a photon couples
to the intermediate pion-baryon states,
must also be included additionally to the valence
quark contributions~\cite{Burkert04,NDelta,NDeltaD,Diaz07a,SatoLee,Kamalov01,Pascalutsa07}.
Thus, the $\gamma N \to \Delta(1232)$ transition
form factor $G_M^\ast$,
may be split into two contributions,
\be
G_M^\ast(Q^2)= G_M^b(Q^2) + G_M^\pi(Q^2),
\label{eqGM}
\ee
where $G_M^b$ and $G_M^\pi$ represent the contributions
from the quark core (also will be denoted by ``bare''),
and those from the pion cloud, respectively.
The above separation is justified if
the pion is created by the baryon of a
three-quark system, and not by a single quark inside the baryon.
Also, according to the chiral perturbation
theory, heavy meson loops are suppressed, and the processes
with one pion loop are dominant in the low
$Q^2$ region~\cite{Gail06,Pascalutsa06a}.
Thus, we restrict the pion-baryon
intermediate states to the lowest order, namely,
"one pion in the air" in the following.

With the decomposition of Eq.~(\ref{eqGM}), we
can separate the short-range contributions
that are sensitive to the quark structure ($G_M^b$)~\cite{Diaz07a},
and those of the long-range which depend on
the pion cloud ($G_M^\pi$).
We note that the same decomposition Eq.~(\ref{eqGM}) was
also applied in several
works~\cite{Diaz07a,SatoLee,Kamalov01,Dong99,Alberto04,Chen08}.

To estimate the quark core contributions $G_M^b$,
one needs a microscopic quark model of baryons.
As for the pion cloud contributions $G_M^\pi$,
one can use a long-range effective dynamics in the low $Q^2$ region
based on chiral symmetry.
According to chiral perturbation theory
the two regimes cannot in general
be disentangled~\cite{ChPT}.
That separation is possible only
in a specific formalism, provided that
the scale of the quark core is defined.

In the covariant spectator quark model
$G_M^b(Q^2)$ was calculated for the
$\gamma N \to \Delta(1232)$ transition
by the processes which a photon directly
couples to the constituent
quarks~\cite{NDelta}.
Here, the overlap integral between the
nucleon and $\Delta$ scalar wave functions
played an important role~\cite{NDelta},
as will be also discussed in Sec.~\ref{secValence}.
The model for the $\Delta(1232)$ structure was calibrated
by the core contributions of the Sato-Lee
model by switching off the pion cloud effects~\cite{Diaz07a}.
Furthermore, the model was successfully able to reproduce the
quenched lattice QCD data \cite{Alexandrou08}
for heavy pions, where the pion cloud effects
are known to be small~\cite{Lattice,LatticeD}.
These facts give us some confidence that the
valence quark contributions of the model are
well under control.

To describe the $\gamma N \to \Delta(1600)$ transition,
we use the same formalism which was successfully applied to
study the $\gamma N \to \Delta(1232)$
transition~\cite{NDelta,NDeltaD,Lattice,LatticeD}
and $\Delta(1232)$ elastic form factors~\cite{DeltaFF0,DeltaFF}.
As the $\Delta(1600)$ resonance shares
many common properties with the $\Delta(1232)$ resonance
such as spin and isospin,
it is reasonable to assume that the $\Delta(1600)$ as the first
radial excitation of the $\Delta(1232)$, and that it can also
be described by a S-state approximation.
Then, one can determine the $\Delta(1600)$ wave function completely
by the orthogonality condition to that of the $\Delta(1232)$.
Using the $\Delta(1600)$ wave function determined in this way
and the nucleon wave function determined
in the previous study~\cite{Nucleon},
we can estimate the valence quark contributions for
the transition magnetic dipole form factor,
$G_M^b$, for the $\gamma N \to \Delta(1600)$ transition.
The detail will be given in next section.

On the other hand, the pion cloud contributions ($G_M^\pi$)
for the $\gamma N \to \Delta(1232)$ transition
in the spectator formalism was estimated
using an effective parametrization~\cite{NDelta}.
The parametrization is consistent with
the pion cloud contributions derived from the dynamical
meson-baryon coupled-channel models of
Sato-Lee~\cite{Diaz07a}, and Mainz-Taipei~\cite{Kamalov01}.
The relative contributions
of the pion cloud ($G_M^\pi/G_M^\ast$)
are simulated by a dipole form factor
which suppresses the pion cloud contributions
in the high $Q^2$ region.

To take account of the pion cloud contributions $G_M^\pi$
in the $\gamma N \to \Delta(1600)$ transition, we must
include more pion-baryon intermediate states
than for the $\gamma N \to \Delta(1232)$ transition.
The detailed discussions concerning the pion cloud
effects are given in Sec.~\ref{secPion}.

In the present work we adopt a hybrid approach
to study the $\gamma N \to \Delta(1600)$ transition,
which was successfully applied for the $\gamma N \to \Delta(1232)$ transition.
This hybrid approach has the following advantages.
It can explain why the bare contributions are insufficient
to describe the electromagnetic transition form factors.
Furthermore, it provides a simple parametrization for $G_M^b$ that
cannot be derived from usual dynamical coupled-channel models.
Finally, it can also incorporate the pion cloud effects
which is justified by dynamical coupled-channel models,
and essential to describe the $\gamma N \to \Delta(1232)$
transition.

\section{Valence quark contributions for the $\gamma N \to
\Delta$ transition}
\label{secValence}

In the covariant spectator quark model
a baryon is described as a system of
three constituent quarks:
one off-mass-shell quark free to interact
with a electromagnetic field,
and two on-shell quarks that act as
an on-shell diquark with a mass $m_D$.
In this formalism~\cite{Nucleon}
the quark-diquark vertex is assumed to be
zero at the singularity point of the three-quark propagator,
and this corresponds to an effective description
of confinement~\cite{Nucleon,Savkli01,Gross06}.
The nucleon and $\Delta(1232)$  states can be well
approximated by a quark-diquark system with zero relative
orbital angular momentum~\cite{Nucleon,NDelta}.
This S-state structure is sufficient to reproduce
the nucleon elastic form factor data~\cite{Nucleon}
and the dominant contributions for the
$\gamma N \to \Delta(1232)$ transition~\cite{NDelta}.
We call this as S-state approach or S-state approximation hereafter.

\subsection{Transition current}

In the spectator quark model
the electromagnetic current for a transition
between an initial state $\Psi_i$ and
a final state $\Psi_f$ is given by,
\be
J^\mu =
3 \sum_{\lambda} \int_k
\overline \Psi_f (P_+,k) j_I^\mu
 \Psi_i (P_-,k),
\ee
where $k$ is the diquark on-shell momentum,
\mbox{$\int_k \equiv \int \sfrac{d^3 k}{2E_D(2\pi)^3}$}
with $E_D$ the diquark on-shell energy,
$P_-$ ($P_+$) is the initial
(final) momentum, $q=P_+-P_-$,
$\lambda$ is the diquark polarization  ($0,\pm 1$)
and $j_I^\mu$ the quark current.
The factor 3 comes from the symmetrization
in the quark flavor
(see Refs.~\cite{Nucleon,Omega,Octet} for details).
In the above, the diquark
polarization and the baryon spin projection
indices are suppressed.

The constituent quark current can be decomposed by,
\be
j_I^\mu= j_1(Q^2)
\left(\gamma^\mu - \frac{\not \! q q^\mu}{q^2}
\right) +
j_2(Q^2)  \frac{i \sigma^{\mu \nu} q_\nu}{2M},
\label{eqjI}
\ee
where $M$ is the nucleon mass.
The Dirac ($j_1$) and Pauli ($j_2$) quark form factors in the above
are also decomposed into the isoscalar and isovector components:
\be
j_i(Q^2)= \frac{1}{6}f_{i+}(Q^2) +   \frac{1}{2}f_{i-}(Q^2) \tau_3,
\hspace{1em} (i=1,2).
\label{eqjI2}
\ee
The quark form factors $f_{i\pm}$
are normalized to $f_{1\pm}(0)=1$ and $f_{2\pm}(0)=\kappa_\pm$
(isoscalar and isovector quark anomalous moments).
Their explicit expressions are given in
Refs.~\cite{Nucleon,NDelta,NDeltaD,Omega,Octet}.

\subsection{Baryon wave functions}

In the S-state approach
the nucleon wave function, $\Psi_N(P,k)$, with $P$
($k$) being the
nucleon (diquark)
momentum, can be written as~\cite{Nucleon},
\be
\Psi_N(P,k)=
\frac{1}{\sqrt{2}}
\left[
\phi_I^0 \phi_S^0 +
\phi_I^1 \phi_S^1
\right] \psi_N(P,k),
\label{eqPSIN}
\ee
where $\phi^{0,1}_{I,S}$ represents isospin ($I$)
or spin ($S$) states corresponding to the
total magnitude of either 0 or 1 in the diquark configuration~\cite{Nucleon}.
In Eq.~(\ref{eqPSIN}), $\psi_N(P,k)$ is the
nucleon scalar wave function to be specified later.

A generic $\Delta$ state (spin 3/2)
with mass $M_\Delta$, is represented
in the S-state approach, as proposed in Ref.~\cite{NDelta},
\be
\Psi_\Delta(P,k) =
- \psi_\Delta(P,k) \tilde \phi_I^1
\varepsilon_P^{\beta \ast}
u_\beta(P),
\label{eqPsiDelta}
\ee
where $P$ ($k$) is the total (diquark) momentum,
$u_\beta$ the Rarita-Schwinger vector-spinor,
$\varepsilon_P$ the polarization vector
in the fixed-axis representation~\cite{FixedAxis},
$\tilde \phi_I^1$ is the isospin state
associated with the isospin-1 diquark
in a spin 3/2 system~\cite{NDelta}
and $\psi_\Delta(P,k)$ is a scalar
wave function also to be specified later.
In the above, both nucleon and $\Delta$ wave functions satisfy
the Dirac equation with respective masses~\cite{Nucleon,NDelta,NDeltaD}.

\subsection{Form factors}

The $\gamma N \to \Delta$ transition,
between a nucleon and a $\Delta$ in S-states,
is characterized by
a magnetic dipole form factor,
\be
G_M^b (Q^2)= \frac{8M}{3\sqrt{3} (M_\Delta+ M)} f_v (Q^2)
{\cal I}_{\Delta N}(Q^2),
\label{eqGMb}
\ee
with
\be
f_v(Q^2)= f_{1-} (Q^2)+ \frac{M_\Delta + M}{2M} f_{2-}(Q^2).
\label{eqfv}
\ee
In Eq.~(\ref{eqGMb}), ${\cal I}_{\Delta N}(Q^2)$ is the overlap integral between
the $\Delta$ and nucleon S-state
scalar wave functions:
\be
{\cal I}_{\Delta N}(Q^2) =
\int_k \psi_\Delta^\ast(P_+,k) \psi_N(P_-,k).
\label{eqInt}
\ee
As in  Eq.~(\ref{eqGMb}) the index $b$ is used to express the "bare".

Eqs.~(\ref{eqGMb})-(\ref{eqInt}) hold
for a generic transition between a spin 1/2 ($N$)
and a spin 3/2 ($\Delta$) baryons described by the S-state approximation.

\subsection{Model for the scalar wave functions}

In the following we use notations,
$\Delta, \Delta^\ast$ and $N^\ast$
for $\Delta(1232), \Delta(1600)$ and $N(1440)$,
respectively, whenever convenient.

To describe the momentum distribution
of the quark-diquark system in a baryon $B$,
we introduce a scalar wave function $\psi_B$,
which depends on the relative angular
momentum and the radial excitation of the system.
As the baryon and the diquark are
on-shell in the covariant spectator model,
the scalar wave function $\psi_B$
can be written as a function of $(P-k)^2$~\cite{Nucleon}.
The dependence on these momenta can be made in term of the
dimensionless variable~\cite{Nucleon},
\be
\chi_B= \frac{(M_B-m_D)^2-(P-k)^2}{M_B m_D},
\label{eqChi}
\ee
where $M_B$ is the baryon mass ($B=N, N^\ast, \Delta, \Delta^\ast$).

The nucleon scalar wave function $\psi_N$
is defined by~\cite{Nucleon},
\be
\psi_N(P,k) = \frac{N_0}{m_D(\beta_1 + \chi_N)(\beta_2 +\chi_N)},
\label{eqPsiN}
\ee
where $\chi_N$ is obtained by inserting $M_B=M$ in Eq.~(\ref{eqChi}),
and $N_0$ the normalization constant \cite{Nucleon}.
In a parametrization where $\beta_2 > \beta_1$,
$\beta_1$ is associated with the long-range
physics, while $\beta_2$ the short-range physics.

As for the $\Delta$, we use the form proposed in Ref.~\cite{NDelta}
based on a S-wave ground state configuration,
\be
\psi_\Delta(P,k) =
\frac{N_1}{m_D(\alpha_1 + \chi_\Delta)(\alpha_2 +\chi_\Delta)^2},
\label{eqPsiD}
\ee
where $N_1$ is the normalization constant,
$\alpha_1$ and $\alpha_2$ are the
parameters which control the momentum ranges
with $\alpha_2  > \alpha_1$.
The $\alpha_1$ is associated with
the long-range physics in the $\Delta$ system \cite{NDelta}.
Note that the difference in the form of $\psi_\Delta$ wave function
from that of the nucleon in Eq.~(\ref{eqPsiN}).
Namely, an extra power in $\psi_\Delta$ exists, and this is preferred
by the magnetic dipole form factor ($G_M^\ast$) data in
the $\gamma N \to \Delta(1232)$
transition~\cite{NDelta,NDeltaD}.
The description of the $\gamma N \to \Delta(1232)$ transition
can be improved with the inclusion of D-states,
which induce also nonzero $G_E^\ast$ and $G_C^\ast$ form factors.
But the inclusion of D-states requires extra parameters~\cite{NDelta,LatticeD}.
In that case (with D-states) the two parameters $\alpha_1,\alpha_2$
are degenerate~\cite{LatticeD}.
In this study we do not include any D-states,
since the S-states are sufficient to describe well
the $\gamma N \to \Delta(1232)$ transition,
with only two parameters in the valence quark sector.

The quality of the present model description
for the $\gamma N \to \Delta(1232)$
can be understood
by comparing with the $G_M^\ast$ data,
or with the helicity amplitudes $A_{1/2}$ and $A_{3/2}$,
which will be shown in Fig.~\ref{figAmp} in Sec.~\ref{secHelAmp}.

For the $\Delta^\ast$ wave function
we assume the same structure as that of the $\Delta$
presented in Eq.~(\ref{eqPsiDelta}),
except for the scalar wave function.
To represent $\Delta^\ast$ as the first radial excitation
of $\Delta$, we write the $\Delta^\ast$ scalar wave function
in the form,
\ba
\psi_{\Delta^\ast}(P,k) &=&
N_2 \frac{\alpha_4 -\chi_{\Delta^\ast}}{(\alpha_3+ \chi_{\Delta^\ast})}
\nonumber \\
& &
\times
\frac{1}{m_D(\alpha_1 + \chi_{\Delta^\ast} )(\alpha_2 +\chi_{\Delta^\ast})^2},
\label{eqPsiDs}
\ea
where $\alpha_3 = \alpha_1$ will be assumed later,
and $\alpha_4$ is a new parameter to be
fixed by the orthogonality condition
between the $\Delta$ and $\Delta^\ast$ states.
The normalization constant $N_2$ will be fixed
by $\int_k |\psi_{\Delta^\ast}|^2=1$ at $Q^2=0$,
similarly to the nucleon and
$\Delta$ cases \cite{Nucleon,NDelta,NDeltaD,DeltaFF0,DeltaFF}.

The extra factor
$\sfrac{\alpha_4 -\chi_R}{(\alpha_3+\chi_R)}$
in Eq.~(\ref{eqPsiDs}),
is motivated by the wave functions
obtained in a harmonic-oscillator potential model
for the three-quark system~\cite{Capstick95,Giannini91,Aznauryan08,Diaz04}.
A similar form was also applied for describing
the Roper resonance~\cite{Roper}.

In the numerical calculation we will use
$\alpha_3=\alpha_1$, assuming that the $\Delta$ and $\Delta^\ast$
are described by the same short-range structure.
The difference between the $\Delta$ and $\Delta^\ast$ systems
appear in the structure, $\alpha_4 -\chi_{\Delta^\ast}$,
scaled by the long-range factor, $\alpha_3+ \chi_{\Delta^\ast}$.
Thus, $\alpha_4$ is the only parameter
characteristic in the $\Delta^\ast$ scalar wave function.
With the scalar wave functions for the $\Delta$ and $\Delta^\ast$
respectively Eqs.~(\ref{eqPsiD}) and~(\ref{eqPsiDs}),
there is no guaranty that the orthogonality condition is satisfied
for an arbitrary value of $\alpha_4$.
The value of $\alpha_4$ will
be determined by imposing the orthogonality condition
for the $\Delta$ and $\Delta^\ast$ states.
This will be explained in next section.
Thus, to write down the $\Delta^\ast$ wave function,
no extra parameter is necessary,
since the parameters $\alpha_1$ and $\alpha_2$
have already been fixed by the $\Delta$
wave function~\cite{NDelta}.

\subsection{Orthogonality condition}

The orthogonality between the $\Delta^\ast$ and $\Delta$
states is ensured if the
overlap integral of the
$\Delta^\ast$ and $\Delta$
wave functions vanishes for $Q^2=0$. This leads to,
\be
\left.
\int_k
\psi_{\Delta^\ast}^\ast(P_{\Delta^\ast},k) \psi_\Delta(P_\Delta,k) \right|_{Q^2=0}=0,
\label{eqOrth}
\ee
the orthogonality between the
scalar wave functions.
In the  $\Delta^\ast$ rest frame the momenta of
the $\Delta$ and $\Delta^\ast$ correspond to $Q^2=0$ are:
\ba
& &
P_{\Delta}=
\left(\frac{M_{\Delta^\ast}^2+M_\Delta^2}{2 M_\Delta},0,0,
-\frac{M_{\Delta^\ast}^2-M_\Delta^2}{2 M_\Delta} \right),
\nonumber \\
& &
P_{\Delta^\ast}=(M_{\Delta^\ast},0,0,0).
\ea
The condition Eq.~(\ref{eqOrth}) may be regarded as the simplest
generalization of the nonrelativistic
orthogonality condition, where equal mass
states are orthogonal when ${\bf q}^2= -Q^2=0$.
See Ref.~\cite{Roper} for more details,
where the same orthogonality condition was applied
for the nucleon and Roper state.

\section{Pion Cloud}
\label{secPion}

\begin{figure}[t]
\includegraphics[width=3.0in]{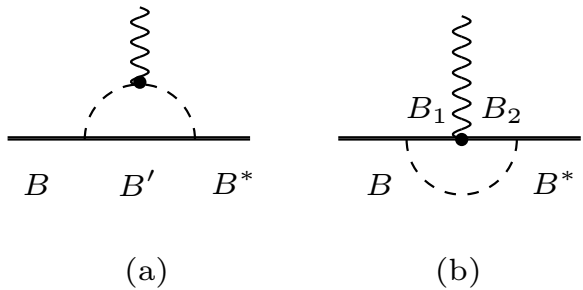}
\caption{\footnotesize
Electromagnetic transition, $\gamma B \to B^\ast$ with
one pion loop (pion cloud) through the intermediate states.}
\label{figPion}
\end{figure}

In the electromagnetic interactions with baryons
there are two main contributions:
a photon couples to an individual quark,
which we will call core or ``bare'',
and a photon couples to the meson-baryon intermediate states,
which we will call meson cloud.
As mentioned already the intermediate states,
where the meson is a pion, are dominant.
In general, contributions from the pion cloud
decreases with increasing $Q^2$ but
can be significant at low $Q^2$.
Among all the processes, one pion ``in the air''
are the most important according to
chiral perturbation theory~\cite{ChPT,Gail06,Pascalutsa06a}.
In the lowest order the dominant processes for
the $\gamma B \to B^\ast$ transition are
shown in Fig.~\ref{figPion}:
(a) a photon couples to the pion,
and (b) a photon couples to the baryon (vertex correction).
The relative contributions of the processes
(a) and (b) are dependent on the systems
and observables in consideration.
We note that the diagram (b) represents
two kinds of interactions:
(b1) a photon interacts with the baryon charge (electric),
and (b2) a photon interacts with the
baryon anomalous magnetic moment (magnetic).

In elastic reactions both contributions (a) and (b1)
must be included consistently for the electric form
factor to satisfy the baryon charge
conservation (see e.g., Ref.~\cite{Octet}).
However, in inelastic transitions like the $\gamma N \to \Delta$ transition,
the dominant contributions are from
the magnetic transition form factor.
When there are two contributions (b1) and (b2),
(b1) is dominant in general.
Also contributions from (a) dominate in general over (b).
This is justified by chiral perturbation theory.
A diagram with two baryon propagators with one pion loop
is suppressed compared to a diagram with two pion propagators~\cite{Arndt04}.
Furthermore, in the study of the octet magnetic moments
with the same spectator formalism, it was indeed
found that the contributions
from the diagram (a) are dominant~\cite{Octet}.
The contributions from the diagram (b) amount to at most
9\% of the diagram (a) for the magnetic moments
except for the $\Lambda$ baryon case\footnote{
The $\Lambda$ baryon case is special, since
it has a small bare magnetic moment.
It has no contributions from the diagrams (a)
but has only from the diagram (b), by the
anomalous coupling of the intermediate state $\Sigma$ baryons,
$\Sigma^+, \Sigma^0$ and $\Sigma^-$~\cite{Octet}.}.
Diagram (a) is also dominant in the decuplet
baryon magnetic moments~\cite{Cloet03}.

Thus, in the present covariant spectator formalism, we can assume
the diagram (a) is dominant for the magnetic form factor
due to the pion cloud, and may neglect the diagram (b)
within an ambiguity of about a 10\%.
Then, we will only focus on the processes
represented by the diagram (a), a photon couples to the pion.

To describe the effect of the pion cloud in the
$\gamma N\to \Delta$ and
$\gamma N\to \Delta^\ast$ transitions, one needs
a microscopic description for the pion-baryon
interactions as well as the photon-pion interactions.
For this, treating the pion as a pointlike particle,
we use the formalism of the cloudy bag model
(CBM)~\cite{Thomas81,Thomas84}.
In CBM, pion couples to a baryon, not to a quark nor
exchanged among the quarks inside the baryon.
This is exactly the same approach as that of
the covariant spectator constituent quark model
used in the present study\footnote{Note that in the spectator
constituent quark model the processes where the pion
is created and absorbed by the same quark are already
included in the constituent quark structure
through the quark electromagnetic form factors.}.
CBM is particularly useful to describe the
pion cloud dressing. For the typical
bag radius the one pion 'in the air' processes
are dominant and the interaction can be
treated perturbatively~\cite{Dodd81,Thomas83,Thomas07}.
We can obtain various pion-baryon coupling constant ratios,
and carry out intermediate state spin and isospin sums
by the formalism based on CBM.
It provides a systematic  method to calculate these
ingredients based on a SU(6) quark model.
Thus, the amplitudes associated with the diagram
(a), represented in terms of the coupling constants,
a coefficient comes from the intermediate spin
and isospin sums, and a scalar integral involving
the quark wave functions, all can be estimated based
on the formalism of CBM.
In the end we replace the respective contributions
due to various intermediate states by an effective
covariant parametrization.
The formalism can be used to relate the pion cloud
contributions associated with different
pion-baryon intermediate states.
We note that the coupling constants used in the present study
are not obtained from CBM, but are calculated from
the decay branches of the resonances using effective Lagrangians.
Only the relevant coupling constant ratios are calculated
based on the CBM formalism.

For the $\gamma N \to \Delta(1232)$ reaction, pion cloud can  
contribute for the form factors $G_M^\ast$, $G_E^ \ast$ and $G_C^\ast$ 
in the spectator quark model~\cite{NDeltaD,LatticeD}.
Although the pion cloud contributions for
the quadrupole form factors can be appreciable 
compared to the valence quark contributions
(from D-states), these contributions are all small compared 
to $G_M^\ast$~\cite{NDeltaD,LatticeD}.
Thus, we consider only the pion cloud
contributions for the magnetic dipole form factor $G_M^\ast$
for the $\gamma N \to \Delta(1232)$ 
and $\gamma N \to \Delta(1600)$ reactions.

In the following we first discuss the pion cloud
contributions for the $\gamma N\to \Delta$ transition,
and then discuss the $\gamma N\to \Delta^\ast$ transition.
Finally, we make a connection for the pion cloud contributions
between the two transitions.

\subsection{$\gamma N \to \Delta(1232)$ transition}

In the description of the $\gamma N \to \Delta$ transition,
the pion cloud contributions for $G_M^\ast$
can be simulated by the
parametrization~\cite{NDelta,NDeltaD},
\be
G_M^\pi (Q^2)= \lambda_\pi \left(
\frac{\Lambda_\pi^2}{\Lambda_\pi^2+ Q^2}
\right)^2 (3 G_D),
\label{eqGMpi}
\ee
where $G_D=(1+ Q^2/0.71)^{-2}$ ($Q^2$ in GeV$^2$)
is the nucleon dipole form factor.
Here, the parameter $\lambda_\pi$ gives
the pion cloud contribution strength,
and $\Lambda_\pi$
is a momentum cutoff parameter.
The simple parametrization of Eq.~(\ref{eqGMpi})
can describe well the main feature of the pion cloud contributions
consistently with the more sophisticated
dynamical coupled-channel models~\cite{Diaz07a,Kamalov01}.
With the parametrization Eq.~(\ref{eqGMpi})
we get a significant contribution
for the pion cloud contributions near $Q^2=0$
(46.4\% at $Q^2=0$, using the parametrization of Ref.~\cite{NDelta}),
and a fast falloff with increasing $Q^2$.
The falloff of the pion cloud contributions is
controlled by $\Lambda_\pi^2$ ($\simeq 1.22$ GeV$^2$)~\cite{NDelta}.
This consistently leads to the dominance of the valence quark contributions
in the region $Q^2 > 3$ GeV$^2$.

Using the CBM~\cite{Thomas83,Thomas84,CBM1,CBM2,CBM3}
framework\footnote{
In our notation $f_{\pi N \Delta}$ corresponds to $f_{\Delta N}$
in CBM~\cite{Thomas83,Thomas84}, }
we can write down the strength of the photon-pion coupling
diagram (a) in Fig.~\ref{figPion},
in the low $Q^2$ region as follows,
\ba
\lambda_\pi &=&
- 2 \sqrt{\frac{2}{3}}
{\cal K} f_{\pi N N} f_{\pi N \Delta}
\hat {\cal C}_{N \Delta} \nonumber \\
& &- 10 \sqrt{\frac{2}{3}}
{\cal K} f_{\pi N \Delta} f_{\pi \Delta \Delta}
\hat {\cal C}_{\Delta \Delta},
\label{eqLpiN}
\ea
where ${\cal K}$ is a generic constant
associated with the interaction and the angular integration.
The factor  $\hat {\cal C}_{B B'}$ represents the ratio,
\be
\hat {\cal C}_{B B'}=\frac{{\cal C}_{B B'}(Q^2)}{{\cal C}_{N \Delta}(Q^2)},
\label{eqIntRatio}
\ee
where ${\cal C}_{B B'}(Q^2)$ is a scalar integral
corresponds to the diagram with the intermediate baryon $B$ and
the final baryon $B^\prime$ states, and ${\cal C}_{N \Delta}(Q^2)$
is the case of $B=N$ and $B^\prime=\Delta$.
The integrals ${\cal C}_{BB'}$ in CBM
depends on the pion-baryon form factor~\cite{Thomas81,Thomas84,Thomas83}.

In Eq.~(\ref{eqLpiN}) the first and the second terms correspond
respectively to the intermediate $N$ and $\Delta$ states
for the diagram (a) in Fig.~\ref{figPion}.
In the ratio in Eq.~(\ref{eqIntRatio}) we expect that the $Q^2$ dependence
largely cancels out, and may regard the ratio as a constant
in the low $Q^2$ region, where the pion cloud is dominant.

The coupling constant $f_{\pi B B'}$ may be calculated
using effective Lagrangians.
In the present study, we use the relative strength to
$f_{\pi N N}=1\, (f_{\pi N N}^2/4\pi=0.08)$,
since what matters is the relative sign and ratio
to the $f_{\pi N N}$, as will be discussed later.

\subsection{$\gamma N \to \Delta(1600)$ transition}

To extend the description of
the pion cloud contributions
from the $\gamma N \to \Delta(1232)$ transition to the
$\gamma N \to \Delta(1600)$ transition,
we include the dominant intermediate states.
In the processes with intermediate baryon state $B$,
$\gamma N \to \pi B  \to \Delta^\ast$,
we include the intermediate states,
$\pi N $, $\pi N(1440)$ and  $\pi \Delta$
as observed by the $\Delta(1600)$ decay~\cite{PDG}.
In addition, the $\pi \Delta(1600)$ intermediate state
is also included.
The strength of the pion cloud contributions from these
intermediate states can be calculated based on the
CBM formalism.
In the following we denote the contributions
from the processes,  $\gamma N \to \pi B \to \Delta(1600)$ with
($B=N,\Delta,N(1440),\Delta(1600)$), by $\lambda_\pi^B$.
The explicit expressions are given by,
\ba
\lambda_\pi^N &=&
- 2 \sqrt{\frac{2}{3}}
{\cal K} f_{\pi N N} f_{\pi N \Delta^\ast}
\hat {\cal C}_{N \Delta^\ast}, \nonumber \\
\lambda_\pi^{N^\ast} &=&
- 2 \sqrt{\frac{2}{3}}
{\cal K} f_{\pi N N^\ast} f_{\pi N^\ast \Delta^\ast}
\hat {\cal C}_{N^\ast \Delta^\ast}, \nonumber \\
\lambda_\pi^\Delta &=&
- 10 \sqrt{\frac{2}{3}}
{\cal K} f_{\pi N \Delta} f_{\pi \Delta \Delta^\ast}
\hat {\cal C}_{\Delta \Delta^\ast}, \nonumber \\
\lambda_\pi^{\Delta^\ast} &=&
- 10 \sqrt{\frac{2}{3}}
{\cal K} f_{\pi N \Delta^\ast} f_{\pi \Delta^\ast \Delta^\ast}
\hat {\cal C}_{\Delta^\ast \Delta^\ast},
\label{eqLambdaP}
\ea
where, $\hat {\cal C}_{B B^\prime}$ is
defined by Eq.~(\ref{eqIntRatio}).

With this procedure, we have reduced the estimate
of the pion cloud contributions
for the $\gamma N \to \Delta(1600)$
to the evaluation of the factor,
\be
\lambda_\pi^\prime = \lambda_\pi^N + \lambda_\pi^{N^\ast} +
\lambda_\pi^{\Delta} + \lambda_\pi^{\Delta^\ast}.
\label{eqLambdaPi}
\ee
Because of the similarity between the
$\gamma N \to \pi B \to \Delta(1600)$
and $\gamma N \to \pi B \to \Delta(1232)$
processes, one can expect that $\lambda_\pi^\prime$
can also be well approximated by a constant,
and the pion cloud contributions can be parameterized by
the same form as that for the $\gamma N \to \Delta(1232)$ transition.
Thus, as in Eq.~(\ref{eqGMpi}), pion cloud contributions for
the $G_M^\ast$ form factor
in the $\gamma N \to \pi B \to \Delta(1600)$ transition
may be given by,
\be
G_M^\pi(Q^2)=
\lambda_\pi^\prime \left(
\frac{\Lambda_\pi^2}{\Lambda_\pi^2+ Q^2}
\right)^2 (3 G_D).
\label{eqGMpi2}
\ee

\subsection{Relation for the pion could contributions
between the $\gamma N \to \Delta(1232)$
and $\gamma N \to \Delta(1600)$ transitions}

It may be a little crude, but as an exploratory study, we neglect
the mass differences of the baryons involved,
and using the formalism based on
CBM~\cite{Thomas83,Thomas84,CBM1,CBM2,CBM3}.
As a consequence we can write
$\hat {\cal C}_{\Delta \Delta}= \hat {\cal C}_{N \Delta}$.
Using the relation from CBM,
$f_{\pi N N} = f_{\pi \Delta \Delta}$,
together with the value $f_{\pi N N} = 1$,
we can rewrite Eq.~(\ref{eqLpiN}) as,
\be
\lambda_\pi =
- 12 \sqrt{\frac{2}{3}}
{\cal K} f_{\pi N \Delta}
\hat {\cal C}_{N \Delta}.
\label{eqLpiN3}
\ee
In the above, the index $N \Delta$ in $\hat {\cal C}_{N \Delta}$
is explicit for a reminder but
note that $\hat{{\cal C}}_{N \Delta}=\hat{{\cal C}}_{NN}$.
With the same approximation for Eq.~(\ref{eqLambdaP}), namely,
$\hat {\cal C}_{B B'} \to \hat{\cal C}_{N \Delta}$,
and together with the result of Eq.~(\ref{eqLpiN3}),
we get the ratios:
\ba
\frac{\lambda_\pi^N}{\lambda_\pi}
 &=&
\frac{1}{6} \frac{f_{\pi N \Delta^\ast}}{ f_{\pi N \Delta}}, \nonumber \\
\frac{\lambda_\pi^{N^\ast}}{\lambda_\pi}
 &=&
\frac{1}{6} f_{\pi N N^\ast}
\frac{f_{\pi N \Delta^\ast}}{ f_{\pi N \Delta}}, \nonumber \\
\frac{\lambda_\pi^{\Delta}}{\lambda_\pi}
 &=&
\frac{5}{6} f_{\pi N \Delta}
\frac{f_{\pi \Delta \Delta^\ast}}{ f_{\pi N \Delta}}
= \frac{5}{6} f_{\pi \Delta \Delta^\ast},
\nonumber \\
\frac{\lambda_\pi^{\Delta^\ast}}{\lambda_\pi}
 &=&
\frac{5}{6}
\frac{f_{\pi N \Delta^\ast}}{ f_{\pi N \Delta}}.
\label{eqLambdaX}
\ea

The coupling constants, $f_{\pi B B'}$, can be
calculated from the $B' \to \pi B$ branching
ratios with some effective Lagrangians at the hadronic level.
This will be discussed in next section.

\subsection{Estimates of the coupling constants $f_{\pi B B^\prime}$}

\begin{table}[t]
\begin{center}
\begin{tabular}{lccc}
\hline
\hline
Decay &$\Gamma$ (MeV) &BR & $f_{\pi N B^\prime}$\\
\hline
$N(1440) \to \pi N $             &300$\pm$100 &0.706$\pm$0.014 & 0.367$\pm$0.061 \\
$\Delta \to \pi N$              &118$\pm$2 &1.00 & 2.160$\pm$0.018 \\
$\Delta(1600) \to \pi N$        &350$\pm$100 &0.153$\pm$0.019 &
0.477$\pm$0.074 \\
\hline
Decay & $\Gamma$ (MeV) &BR & $f_{\pi B \Delta^\ast}$\\
\hline
$\Delta(1600) \to \pi \Delta $   &350$\pm$100 &0.590$\pm$0.100 & 0.653$\pm$0.108\\
$\Delta(1600) \to \pi N(1440) $  &350$\pm$100 &0.130$\pm$0.040 & 6.330$\pm$1.329 \\
\hline
\hline
\end{tabular}
\end{center}
\caption{
Data for resonances from PDG~\cite{PDG},
and the coupling constants calculated.
For $f_{\pi N N}$, we use $f_{\pi N N} = 1$
($f_{\pi N N}^2/4\pi = 0.08$), and the relation based on
CBM~\cite{Thomas81,Thomas84,CBM1,CBM2,CBM3},
$f_{\pi N N}=f_{\pi \Delta \Delta}=f_{\pi \Delta^\ast \Delta^\ast}$.
For the branching ratios, we take an average weighted by the error
of the selected results from PDG.
Errors in coupling constants are estimated using gaussian quadrature.}
\label{tablefNB}
\end{table}

The interaction Lagrangians and
definitions of the coupling constants $\pi B B'$
relevant in this study, are given in Appendix~\ref{decayrates}.
Based on these interaction Lagrangians and
decay rate expressions, we obtain the absolute values of
the coupling constants.
The data used for the calculation, extracted
form Particle Data Group~\cite{PDG},
are summarized in Table~\ref{tablefNB}.
To determine the relative signs for the coupling
constants we follow some
quark models~\cite{Aznauryan07,Thomas81,Thomas84,Riska01}.
For $\pi NN$ constant, we use
the positive value $f_{\pi N N} = 1$
(or  $f^2_{\pi N N}/4\pi = 0.08$), since what matters
is the relative sign and strength to $f_{\pi N N}$.
For the signs of the coupling constants,
$f_{\pi N N(1440)}$, $f_{\pi N \Delta}$ and $f_{\pi N \Delta(1600)}$,
we take the same sign as that of the $f_{\pi N N}$
as suggested by the quark model results~\cite{Riska01}.
For a detailed discussion about the sign of
$f_{\pi N N(1440)}$ see also Ref.~\cite{Aznauryan07}.
Then, the relative signs undetermined are those
for $f_{\pi \Delta \Delta(1600)}$ and $f_{\pi N(1440) \Delta(1600)}$.
Since $\Delta$ and $\Delta(1600)$ differ only
in radial excitations (and thus mass),
we assume the same relative sign
for these coupling constants.  The same argument also holds for
the case where $N$ is replaced by $N(1440)$.
As a result all the coupling
constants relevant in this study
are assigned to the same sign as
that of the $f_{\pi NN}$.
The coupling constants calculated in these manners,
are presented in Table~\ref{tablefNB}.
The values obtained in the present study
are similar to those obtained in Refs.~\cite{Julich,Riska01}.

\section{Helicity amplitudes and form factors}
\label{secHelAmp}

\begin{figure*}[htb]
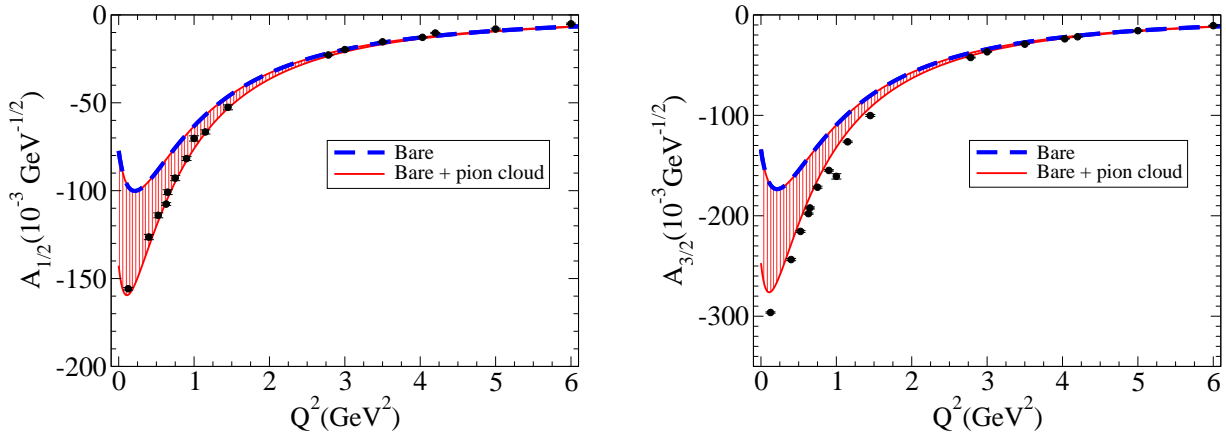

\vspace{.4cm}
\centerline{
\mbox{
\includegraphics[width=3.0in]{A12b}  \hspace{.7cm}
\includegraphics[width=3.0in]{A32b}}}
\caption{\footnotesize
Helicity amplitudes calculated for the
$\gamma N \to \Delta(1232)$ transition in
the S-state approach,
with and without the pion cloud contributions.
Data are taken from the MAID analysis~\cite{Drechsel07}.
}
\label{figAmp}
\end{figure*}

As already mentioned, the description of
the $\gamma N \to \Delta$ transition
is characterized by the three independent multipole
form factors, $G_M^\ast$, $G_E^\ast$ and $G_C^\ast$~\cite{Jones73}.
These form factors are exclusive functions
of the four-momentum transfer squared $Q^2$
and frame independent.
The physical properties of the $\gamma N \to \Delta$ transition
are usually expressed in terms of the transition
amplitudes in a particular frame.
As there are amplitudes associated with any
photon polarization including the longitudinal polarization,
there are three independent transition
amplitudes, $A_{1/2}$, $A_{3/2}$
and $S_{1/2}$~\cite{Aznauryan07,Aznauryan08}.
The helicity amplitudes for the transitions,
$\gamma N \to \Delta$ or $\gamma N \to \Delta(1600)$
at the final particle rest frame, can be related with the form
factors by~\cite{Capstick95}:
\ba
G_M^\ast(Q^2) = -F(Q^2) \left[
\sqrt{3} A_{3/2}(Q^2) + A_{1/2}(Q^2) \right],
\label{eqGMs}\\
G_E^\ast(Q^2) = -F(Q^2) \left[
\frac{1}{\sqrt{3}} A_{3/2}(Q^2) - A_{1/2}(Q^2) \right].
\label{eqGEs}
\ea
In the above the factor $F(Q^2)$ is given by,
\be
F(Q^2) = \frac{1}{e}
\sqrt{\frac{M(M_\Delta^2-M^2)}{2
\left[(M_\Delta-M)^2 + Q^2\right]}}
\frac{2M}{M_\Delta+M},
\ee
where $e=\sqrt{4 \pi \alpha}$ is the
magnitude of the electron charge, with $\alpha = 1/137.036$
the fine-structure constant.
There is an extra relation between the transverse amplitude $S_{1/2}$
and $G_C^\ast$, but we omit it since it is irrelevant in the present study.

In a model with an S-state approach for the nucleon and $\Delta$
and the pion cloud contributes only for $G_M^\ast$, 
$G_M^\ast$ is dominant and one has 
$G_E^\ast \equiv 0$, $G_C^\ast \equiv 0$~\cite{NDelta,NDeltaD}. 
In these conditions with $G_E^\ast (Q^2)= 0$ for an arbitrary $Q^2$,
we get:
\ba
& &
A_{3/2}(Q^2)= -\frac{\sqrt{3}}{2F (Q^2)} G_M^\ast(Q^2),\\
& &
A_{1/2}(Q^2)=  -\frac{1}{2F(Q^2)} G_M^\ast (Q^2).
\ea
Thus, we can write the helicity amplitudes $A_{1/2}$ and $A_{3/2}$
in terms of $G_M^\ast$ for an arbitrary $Q^2$.
As for the transverse amplitude $S_{1/2}$,
which is proportional to $G_C^\ast$, one has $S_{1/2}\equiv 0$.

The helicity amplitudes $A_{1/2}$ and $A_{3/2}$
for the $\gamma N \to \Delta(1232)$ in the
S-state approach,
as well as the contributions from the quark core,
are presented in Fig.~\ref{figAmp},
From the figure, one can see that
the S-state approximation plus
pion cloud dressing (for $G_M^\ast$)
reproduces well the data
for $\gamma N \to \Delta(1232)$ transition.
Encouraged by this, we will use the same approximation
for the $\gamma N \to \Delta(1600)$ transition.
The S-state approach will be tested in next section.

\section{Results}
\label{secResults}

\begin{table}[t]
\begin{center}
\begin{tabular}{lccc}
\hline
\hline
$\gamma N \to \pi B  \to \Delta(1600)$
& $f_{\pi N B}$ & $f_{\pi B \Delta(1660)}$ & $\lambda^B_\pi/\lambda_\pi$   \\
\hline
$\gamma N \to \pi N \to \Delta(1600)$       & 1.000 &0.477 & 0.0368$\pm$0.0057\\
$\gamma N \to \pi N(1440)\to \Delta(1600)$ & 0.361 &6.330 & 0.1791$\pm$0.0481\\
$\gamma N \to \pi \Delta \to \Delta(1600)$  & 2.160 &0.653 & 0.5441$\pm$0.0904\\
$\gamma N \to \pi \Delta(1600)\to \Delta(1600)$ & 0.477  & 1.000   &
0.1842$\pm$0.0287   \\
\hline
Total & & & 0.9442$\pm$0.1065\\
\hline
\hline
\end{tabular}
\end{center}
\caption{
Results for the coupling constants and $\lambda_\pi^B$
($B=N,N(1440),\Delta,\Delta(1600)$.
The total contribution of the pion
cloud is given by the sum of $\lambda_\pi^B$,
which amounts to $0.9442 \lambda_\pi$,
where $\lambda_\pi=0.464$ \cite{NDelta}.
The uncertainty in the final result (Total)
is obtained by adding the
errors in gaussian quadrature. }
\label{tablePionCloud}
\end{table}

In this section we present numerical results
for the form factors and helicity amplitudes.
Because of the approximation used in this exploratory study,
it holds that $G_E^\ast=0$ and $G_C^\ast=0$ for all $Q^2$.
Thus, we have nonzero results only for $G_M^\ast$.
The electric (E2) and Coulomb (C2) quadrupole form
factors both vanish, as well as the ratios, E2/M1 and C2/M1.
We start by the case $Q^2=0$ and compare
the results with the available experimental data.
Next we discuss the $Q^2$ dependence of the form
factors and make some predictions.

The experimental information
for the $\gamma N \to \Delta(1600)$
transition is restricted to the helicity amplitudes
$A_{1/2}$ and $A_{3/2}$ measured in the
$\Delta(1600)$ rest frame at the photon point ($Q^2=0$).
This is collected in Ref.~\cite{PDG} by PDG.
Particle Data Group selected three results:
Awaji, Crawford~\cite{Crawford83} and
Arndt~\cite{Arndt96}.
The result of Awaji has a large uncertainty.
The results from PDG are presented in Table~\ref{tabGM},
together with the result calculated for $G_M^\ast(0)$ and $G_E^\ast(0)$
by Eqs.~(\ref{eqGMs}) and~(\ref{eqGEs}).

\subsection{Analysis in the limit $Q^2=0$}

We discuss first the contributions from the
valence quarks ($G_M^b$).
In the S-state approach $G_M^b$ is given by Eq.~(\ref{eqGMb}).
With the scalar wave function Eq.~(\ref{eqPsiD}),
and the mass $M_\Delta$,
we get the result for the $\gamma N \to \Delta(1232)$ transition.
Similarly, with the scalar wave function Eq.~(\ref{eqPsiDs}),
and $M_\Delta$ replaced by $M_{\Delta^\ast}$, we can get
the result for the $\gamma N \to \Delta(1600)$ transition.
Note that it is the scalar wave function
$\psi_\Delta$ or $\psi_{\Delta^\ast}$
that characterizes the radial state
(ground state or first radial excited state).
We adopt model II in Ref.~\cite{NDelta}: $\alpha_1=0.290$
and $\alpha_2=0.393$. The normalization constant is $N_1=2.95$.
The $\Delta^\ast$ wave function is determined by the
scalar wave function Eq.~(\ref{eqPsiDs}) with
$\alpha_3= \alpha_1$, where $\alpha_1$ is
the parameter associated with the long-range scale.
The unknown parameter $\alpha_4$ is determined
by the orthogonality condition Eq.~(\ref{eqOrth}),
which gives $\alpha_4= -0.0353$.
The corresponding normalization constant for the
$\Delta^\ast$ scalar wave function is $N_2 = 7.27$.
The parameters associated with the nucleon scalar
wave function are given by model II of Ref.~\cite{Nucleon}.
With these parameters fixed, the overlap integral between the
$\Delta^\ast$ and nucleon scalar wave function at $Q^2=0$
given by Eq.~(\ref{eqInt})
with $\Delta$ replaced by $\Delta^\ast$, is calculated:
\be
{\cal I}_{\Delta^\ast N}(0)= -0.564.
\ee
Then, the contributions
from the valence quarks (bare) for the magnetic dipole
form factor at $Q^2=0$ of Eq.~(\ref{eqGMb}) result to,
\be
G_M^b(0)=-1.113.
\label{eqGMmodel}
\ee

Thus,
the valence quark core contributions
underestimate largely
the experimental values and differ in sign
from the data shown in Table~\ref{tabGM}.
We recall that in the present approach $G_E^\ast \equiv 0$.

Now, we turn our discussion to the pion cloud contributions.
The pion cloud contributions for the transition magnetic form factor
is estimated by Eq.~(\ref{eqGMpi2}).
The coefficient $\lambda_\pi^\prime$ can be
obtained by Eqs.~(\ref{eqLambdaPi}) and~(\ref{eqLambdaX}).
This includes contributions from the dominant intermediate
baryon states, $N,N(1440),\Delta,\Delta(1600)$.
These contributions depend on the $\pi B B'$ coupling constants.
The coupling constants calculated from experimental data~\cite{PDG}
are presented in Table~\ref{tablefNB}.
Using these values we calculate $\lambda_\pi^\prime$,
and each intermediate state contribution is
listed in Table~\ref{tablePionCloud}.
Then, we get the total contribution for the pion cloud,
relative to those of the $\gamma N \to \Delta(1232)$:
\be
\frac{\lambda_\pi^\prime}{\lambda_\pi}=
0.944\pm0.107.
\ee
Once $\lambda_\pi^\prime$ is fixed, the pion cloud
contributions for $G_M^\pi(0)$ is determined by Eq.~(\ref{eqGMpi2})
with $3 G_D(0)=3$:
\ba
G_M^\pi(0)&=& \frac{\lambda_\pi^\prime}{\lambda_\pi} (3 \lambda_\pi)
\nonumber \\
&=& 1.314\pm0.148.
\label{eqGMpiM}
\ea

Adding the valence quark
contributions and the pion cloud contributions,
Eqs.~(\ref{eqGMmodel}) and~(\ref{eqGMpiM}), respectively,
we get,
\be
G_M^\ast(0)=0.202\pm0.131.
\ee
This result is compared with experimental data in Table~\ref{tabGM}.
The more accurate data available~\cite{Crawford83,Arndt96}
supports the S-state approximation,
and the consequent $G_M^\ast$ dominance.
The corresponding results for the helicity amplitudes,
$A_{1/2}(0)$ and $A_{3/2}(0)$ are also presented
in Table~\ref{tabGM}.

\begin{table*}
\begin{center}
\begin{tabular}{l l c c c}
\hline
\hline
  & $A_{1/2}(0)(\mbox{GeV}^{-1/2})$ &
$A_{3/2}(0)(\mbox{GeV}^{-1/2})$ & $G_M^\ast(0)$  & $G_E^\ast(0)$ \\
\hline
Awaji 1981~\cite{PDG} & $-0.046\pm0.013$  & $+0.025\pm0.031$ & $0.009\pm0.181$
& $-0.198\pm0.073$ \\
Crawford 1983~\cite{Crawford83} & $-0.039\pm0.030$  & $-0.013\pm0.014$ & $0.202\pm0.127$
& $-0.103\pm0.102$ \\
Arndt 1996~\cite{Arndt96}
& $-0.018\pm0.015$  & $-0.025\pm0.015$ & $0.201\pm0.098$
& $-0.012\pm0.057$ \\
\hline
Model & $-0.0154\pm0.0113$  & $-0.0266\pm0.0196$ & $0.202\pm0.148$
& 0.000 \\
\hline
\hline
\end{tabular}
\end{center}
\caption{Results at $Q^2=0$ compared with the
selected data from PDG~\cite{PDG}.
$G_E^\ast=0$ is the consequence of the S-state approximation.}
\label{tabGM}
\end{table*}

\subsection{$Q^2$ dependence of $G_M^\ast$}

\begin{figure}[t]
\vspace{.2cm}
\centerline{
\mbox{
\includegraphics[width=2.9in,angle=0]{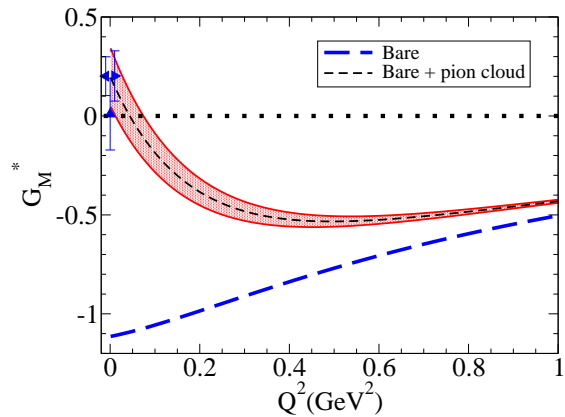} }}
\caption{\footnotesize{
$\gamma N \to \Delta(1600)$ magnetic dipole form factor.}}
\label{figGM}
\end{figure}

In our model the magnetic dipole form factor
$G_M^\ast(Q^2)$ is given by
the sum of
$G_M^b(Q^2)$ and $G_M^\pi(Q^2)$.
The valence quark contributions are
given by Eq.~(\ref{eqGMb}), which includes
the isovector factor $f_v(Q^2)$ and
the $Q^2$ dependent overlap integral ${\cal I}_{\Delta^\ast N}(Q^2)$
between the $\Delta^\ast$ and nucleon scalar wave
functions [see Eqs.~(\ref{eqfv})-(\ref{eqInt})].
The $Q^2$ dependence of $G_M^\ast$ is shown
in Fig.~\ref{figGM}.
As for the pion cloud contributions $G_M^\pi$,
these are determined by Eq.~(\ref{eqGMpi2}),
once the coefficient $\lambda_\pi^\prime$ is known.
The band in Fig.~\ref{figGM}
shows the uncertainty in the
estimate of the coupling $f_{\pi B B'}$
from the data listed in Table~\ref{tablefNB}.

\begin{figure}[t]
\vspace{.3cm}
\centerline{
\mbox{
\includegraphics[width=2.9in,angle=0]{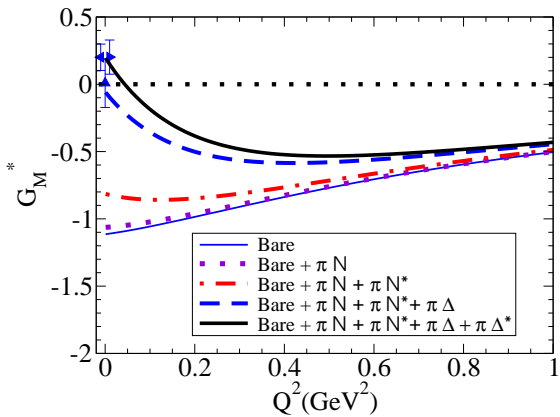} }}
\caption{\footnotesize{
Decomposition of the contributions for the
$\gamma N \to \Delta(1600)$ magnetic dipole form factor.
}}
\label{figGM2}
\end{figure}

\begin{figure}[t]
\vspace{.5cm}
\centerline{
\mbox{
\includegraphics[width=2.9in,angle=0]{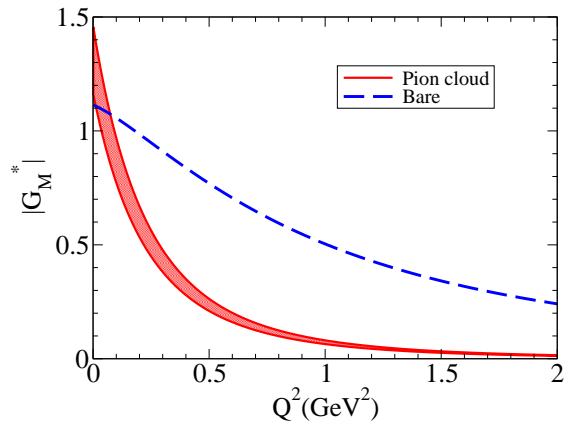} }}
\caption{\footnotesize{
Absolute values of the bare and the pion cloud contributions for the
$\gamma N \to \Delta(1600)$ transition magnetic form factor.}}
\label{figGMmeson}
\end{figure}

Each pion cloud contribution due to the different
intermediate states, $N,N^\ast \Delta,\Delta^\ast$,
is shown in Fig.~\ref{figGM2}, in an accumulative manner.
As the pion cloud contributions from the different
intermediated states are added one by one,
the result for $G_M^\ast(0)$
approaches to the experimental data points accordingly.
In Fig.~\ref{figGM2}, uncertainties in
the pion cloud contributions are not shown for clarity.
In the figure one can see that the $\pi \Delta$
intermediate state gives the dominant contribution.
According to the values in Table~\ref{tablePionCloud},
the $\pi \Delta$ intermediate state contribution is about 48-67\%
of the total pion cloud contribution.
The contributions from the $\pi N^\ast $ and $\pi \Delta^\ast $
intermediate states amount to about 33-44\% of the total
pion cloud contribution.
Figure~\ref{figGM2} shows also a faster falloff
of the pion cloud contributions with increasing $Q^2$,
compared to the $Q^2$ dependence of the quark core.
This can be better seen in Fig.~\ref{figGMmeson},
where absolute values of bare and pion cloud contributions
are compared.
In the same figure one can also see
the pion cloud contributions
are dominant near $Q^2 =0$, while the bare (quark core)
contributions ($G_M^b$) become dominant in the region $Q^2 > 0.5$ GeV$^2$.

\subsection{$Q^2$ dependence of $A_{1/2}$ and $A_{3/2}$}

\begin{figure*}[htb]
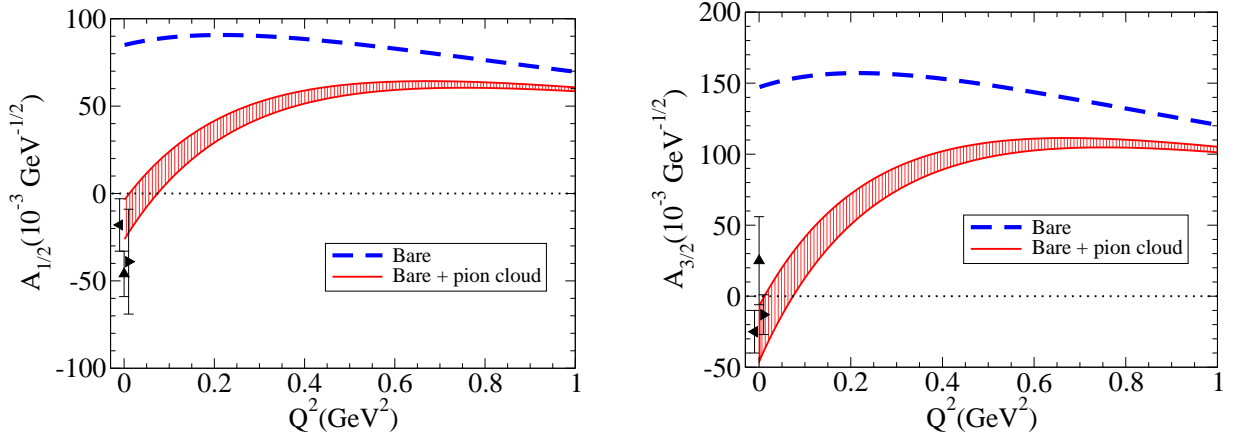

\vspace{.4cm}
\centerline{
\mbox{
\includegraphics[width=3.0in]{A12_1600a}  \hspace{.7cm}
\includegraphics[width=3.0in]{A32_1600a}}}
\caption{\footnotesize
Helicity amplitudes for the
$\gamma N \to \Delta(1232)$ transition, calculated in
the S-state approximation for the $N$ and $\Delta(1600)$.
Data from PDG~\cite{PDG}. See Table \ref{tabGM}}
\label{figAmp1600}
\end{figure*}

$Q^2$ dependence of the helicity amplitudes,
$A_{1/2}(Q^2)$ and $A_{3/2}(Q^2)$,
can be obtained in the $\Delta(1600)$ rest frame.
In the S-state approach discussed
in Sec.~\ref{secHelAmp}, the amplitudes are given
by Eqs.~(\ref{eqGMs}) and~(\ref{eqGEs}).
The results are shown in Fig.~\ref{figAmp1600}.
In the figure the contributions
of the quark core (bare) are also shown.
We predict from Fig.~\ref{figAmp1600} that
$A_{1/2}(Q^2)$ and $A_{3/2}(Q^2)$ become
positive for $Q^2 > 0.1$ GeV$^2$.
This result is consistent with
the estimates made in Ref.~\cite{Capstick95},
which are based on the valence quark structure.
The positive sign in the helicity amplitudes
for $Q^2 > 0.1$ GeV$^2$ is essentially a consequence
of the quark core dominance.

\subsection{Discussion}

Our results for $G_M^\ast(0)$ (central value)
is very close to the experimental data of
Refs.~\cite{Arndt96,Crawford83}.
The result is also consistent with the data of Awaji~\cite{PDG} within the
error bars, but the data are not consistent with
$G_E^\ast \equiv 0$ of the present approach.
However, one should keep in mind that the present results
are based on the approximation of ignoring
the baryon mass differences in the estimate of
the pion cloud contributions,
and on the dominance of the photon-pion coupling diagram
[diagram (a) in Fig.~\ref{figPion}],
which has a 10\% ambiguity.
Unfortunately, we cannot draw more definite conclusions,
since the uncertainty associated with
the pion cloud contributions is $0.148$, which
is comparable with the central value $G_M^\ast(0)=0.202$,
and also relatively large experimental errors exist.
The large uncertainty in our estimate lies mainly in the
$\pi \Delta$ intermediate state, in particular
the coupling constant $f_{\pi \Delta \Delta^\ast}$.
An accurate value of the coupling constant
would reduce the final uncertainty almost
by a factor of two.
A better constraint of the pion cloud contribution
can be achieved once better experimental data become
available associated with the $\Delta^\ast$ decay
to extract $f_{\pi \Delta \Delta^\ast}$.
An alternative may be to use
the coupling constants from an independent model
for the meson-baryon interaction,
where the coupling constants are constrained
by many observables.
At the moment such well constrained coupling constants
associated with the $\Delta(1600)$
are not available.

There is also uncertainty
in the expression for the valence quark contributions $G_M^b$.
In the $\Delta^\ast$ scalar wave function
Eq.~(\ref{eqPsiDs}), there is an extra degree
of freedom associated with the momentum
scale parameter $\alpha_3$, which sets
the scale of the variation of the $\Delta^\ast$
wave function compared to that of the ground state $\Delta$.
As explained in the text, we have fixed $\alpha_3$
by the long-range scale parameter $\alpha_1$
(same short-range structure for $\Delta$ and $\Delta^\ast$).
The choice, $\alpha_3= \alpha_2$, would change
the contributions of the core to $G_M^b(0)=-0.924$,
to be compared with the result we have obtained, $-1.113$.
An alternative method may be
to adjust the parameters by fitting to the data
of the helicity amplitudes or form factors,
once they become available for finite $Q^2$.
However, the advantage of the present approach
to focus on the long-range scale parameter,
has also been proven to be good
in the study of the $\gamma N \to P_{11}(1440)$
transition form factors~\cite{Roper}.

\section{Conclusions}
\label{secConclusions}

In this article, we have studied the $\Delta(1600)$ structure,
and the $\gamma N \to \Delta(1600)$ transition
using a covariant spectator formalism,
with a simplified $G_M^\ast$ dominance model.
As far as the authors are aware,
this is the first
dynamical study for the $\gamma N \to \Delta(1600)$ transition
including both the bare and meson cloud contributions.
The role of the $\Delta(1600)$ resonance
in the meson-baryon coupled-channel models
has not been settled yet.
Thus, theoretical study of this resonance
can be a challenge for many baryon models.
Our result show that solely the contributions
from the quark core to the dominant form factor $G^\ast_M$ at $Q^2=0$,
is negative and far below the existing
experimental data points, which are positive.
However, the explicit inclusion of the
pion cloud contributions, overcome the
negative contributions of the valence quark core
to lead to the positive sign, which is consistent
with the  experimental positive values.
The final result, although it has uncertainties
associated with the coupling constants and
approximations used, is consistent with the experimental data.
Furthermore, the present study may provide a parametrization
for the $\Delta^\ast$ core that can be used in
coupled-channel models.

It will be also very interesting to compare
our estimate of the quark core contributions
with the lattice QCD simulation data.
Such simulations were performed in the past
for the $\gamma N \to \Delta(1232)$ transition~\cite{Alexandrou08}.
Finally, we have predicted the $Q^2$
dependence of the $G^\ast_M(Q^2)$ form factor.
Based on the recent analysis for
the electromagnetic structure
of the $P_{11}(1440)$, $D_{13}(1520)$, $S_{11}(1535)$ and
$S_{11}(1650)$, it is expected that also
$P_{33}(1600)$ will be included
in the multipole analysis in the near future.

The method used to estimate the pion cloud
contributions is based on the processes
that a photon couples directly to the pion,
which may be justified by the previous
study for the octet baryon magnetic moments
in the same covariant spectator quark model,
within about a 10\% error.
Then, based on the cloudy bag model formalism,
we have made a connection for the pion cloud
contributions between the $\gamma N \to \Delta(1232)$
and $\gamma N \to \Delta(1600)$ transitions, by
summing over all the intermediate spin and isospin states.
In this exploratory study, we have
approximated the masses of all the intermediate
state baryons by an average value of the
$N$, $N(1440)$, $\Delta$ and $\Delta(1600)$.
In the future we need to include explicitly
the mass differences and treat the pion-baryon
intermediate states properly.

We can apply the present
valence quark model of the baryon
with meson cloud dressing to other systems.
One possibility is the $P_{11}(1440)$ and $P_{11}(1710)$
resonances, where the meson cloud dressing is expected
to be very important in the small $Q^2$ region,
since $P_{11}(1440)$ is described
as the first radial excitation of the nucleon~\cite{Roper},
and the $P_{11}(1710)$ resonance may also be considered
as the second radial excitation of the nucleon~\cite{Capstick95}.
Another possible application of the model
may be to study the octet to decuplet baryon electromagnetic
transitions, by extending the treatment for the
$\gamma N \to \Delta(1232)$ to the SU(3) sector,
where the meson cloud dressing,
pion in particular, is also expected to
be important~\cite{NDeltaD,LatticeD}.


\vspace{0.3cm}
\noindent
{\bf Acknowledgments:}

\vspace{0.2cm}

The authors would like to thank Y.~Kohyama for providing K.T.\ in the past
the calculation note of CBM which has helped the present study.
The authors also would like to thank B.~Juli\'a-D{\'{i}}az, H.~Kamano,
for helpful discussions.
G.~R.\ would like to thank Franz Gross and the Jefferson Lab Theory Group
for the invitation and hospitality during the
period of February and March in 2010.
G.~R.\ was supported by the Portuguese Funda\c{c}\~ao para
a Ci\^encia e Tecnologia (FCT) under the grant
SFRH/BPD/26886/2006.
This work is also supported partially by the European Union
(HadronPhysics2 project ``Study of strongly interacting matter''),
and partially by Jefferson Science Associates, LLC under
U.~S.~DOE Contract No.~DE-AC05-06OR23177.
Notice: The U.~S.~Government retains a non-exclusive,
paid-up, irrevocable, world-wide license
to publish or reproduce this manuscript for U.~S.~Government purposes.

\appendix

\section{Decay rates and coupling constants associated with resonances}
\label{decayrates}

In this appendix we calculate the coupling constants
$f_{\pi B B'}$ necessary
to estimate the pion cloud contributions
based on the available experimental data~\cite{PDG}.
The necessary coupling constants are,
$f_{\pi N \Delta}$, $f_{\pi N N(1440)}$, $f_{\pi N \Delta(1600)}$,
$f_{\pi \Delta \Delta(1600)}$, and $f_{\pi N(1440) \Delta(1600)}$.
For the other coupling constants, we use
$f_{\pi NN}=1$ ($f_{\pi NN}^2/4\pi=0.08$)
and the relation based on CBM~\cite{Thomas81,Thomas84},
$f_{\pi NN}=f_{\pi\Delta\Delta}=f_{\pi \Delta(1600) \Delta(1600)}$.
First, we present the effective Lagrangian densities
used for the calculation of the coupling constants.

The Lagrangian densities used in the present study are:
\ba
{\cal L}_{\pi N N} &=&
- \frac{f_{\pi N N}}{m_\pi}
\left[\bar{N} \gamma^\mu \gamma_5 \mb{\bm$\tau$}
N \right] \cdot {\partial}_\mu \mb{\bm$\pi$},
\nonumber \\
%
{\cal L}_{\pi N N^\ast} &=&
- \frac{f_{\pi N N^\ast}}{m_\pi}
\left[\bar{N}^\ast \gamma^\mu \gamma_5 \mb{\bm$\tau$}
N \right] \cdot {\partial}_\mu \mb{\bm$\pi$} + h.c.,
\nonumber \\
%
{\cal L}_{\pi N \Delta} &=&
\frac{f_{\pi N \Delta}}{m_\pi} \left[\bar{N}
{\bf T} \Delta^\mu \right] \cdot {\partial}_\mu \mb{\bm$\pi$} + h.c.,
\nonumber \\
%
{\cal L}_{\pi N \Delta^\ast} &=&
\frac{f_{\pi N \Delta^\ast}}{m_\pi} \left[\bar{N}
{\bf T} \Delta^{\ast \mu} \right] \cdot {\partial}_\mu \mb{\bm$\pi$} + h.c.,
\nonumber \\
{\cal L}_{\pi N^\ast \Delta^\ast} &=&
\frac{f_{\pi N^\ast \Delta^\ast}}{m_\pi} \left[\bar{N}^\ast
{\bf T} \Delta^{\ast \mu} \right] \cdot {\partial}_\mu \mb{\bm$\pi$} + h.c.,
\nonumber \\
{\cal L}_{\pi \Delta \Delta^\ast} &=&
- \frac{f_{\pi \Delta \Delta^\ast}}{m_\pi}
\left[\bar{\Delta}^\mu \gamma^\nu \gamma_5
{\bf I} \Delta^\ast_\mu \right] \cdot {\partial}_\nu \mb{\bm$\pi$} + h.c.,
\label{LpiDDs}
\ea
where, \mb{\bm$\tau$} are the Pauli matrices, ${\bf T}$ and ${\bf I}$ are the isospin operators
defined by, $({\bf T})_{M m} \equiv \sum_\mu (1 \mu \frac{1}{2} m| \frac{3}{2} M) \hat{e}^\ast_\mu$
and
$({\bf I})_{M M'} \equiv \frac{\sqrt{15}}{2} \sum_\mu (1 \mu \frac{3}{2} M'| \frac{3}{2} M) \hat{e}^\ast_\mu$,
respectively.
Then, one can calculate decay rates and obtain the necessary coupling constants
associated with the resonances.
Widths, branching ratios, and calculated coupling constants of the resonances
are summarized in Table~\ref{tablefNB}.

Next, we give expressions for decay rates calculated using the
Lagrangian densities Eqs.~(\ref{LpiDDs})
to estimate the coupling constants.

The coupling constants are estimated from the
decay rate expressions (see also Ref.~\cite{Tsushima99}):
\ba
& &\hspace*{-8em}\hspace*{-2em}\Gamma (N(1440) \to \pi N) \nonumber\\
& &\hspace{-9em} = 3 \frac{f_{\pi N N(1440)}^2}{4\pi} \frac{(M_N + M_{N(1440)})^2}{M_\pi^2}
\frac{(E_N - M_N)|\vec{p}|}{M_{N(1440)}},
\label{gammaRopNpi}\\
{\rm with}\hspace{1em} |\vec{p}|
&=& \frac{\lambda^{1/2}(M_{N(1440)}^2,M_N^2,m_\pi^2)}{2M_{N(1440)}},
\nonumber
\ea

\ba
& &\hspace*{-8em}\Gamma (\Delta \to \pi N ) =
\frac{f_{\pi N \Delta}^2}{12\pi m_\pi^2} \frac{(E_N + M_N)|\vec{p}|^3}{M_{\Delta}},
\label{gammaDNpi}\\
{\rm with}\hspace{1em} |\vec{p}|
&=& \frac{\lambda^{1/2}(M_{\Delta}^2,M_N^2,m_\pi^2)}{2M_{\Delta}},
\nonumber
\ea

\ba
\hspace*{-2em}\Gamma (\Delta(1600) \to \pi N ) &=&
\frac{f_{\pi N \Delta(1600)}^2}{12\pi m_\pi^2} \frac{(E_N + M_N)|\vec{p}|^3}{M_{\Delta(1600)}},
\label{gammaDsNpi}\\
{\rm with}\hspace{1em} |\vec{p}|
&=& \frac{\lambda^{1/2}(M_{\Delta(1600)}^2,M_N^2,m_\pi^2)}{2M_{\Delta(1600)}},
\nonumber
\ea

\ba
& &\hspace*{-8em}\Gamma (\Delta(1600) \to \pi N(1440)) \nonumber\\
& &\hspace*{-7.5em} = \frac{f_{\pi N(1440) \Delta(1600)}^2}{12\pi m_\pi^2}
\frac{(E_{N(1440)} + M_{N(1440)})|\vec{p}|^3}{M_{\Delta(1600)}},
\label{gammaDsRoppi}\\
{\rm with}\hspace{1em} |\vec{p}|
&=& \frac{\lambda^{1/2}(M_{\Delta(1600)}^2,M_{N(1440)}^2,m_\pi^2)}{2M_{\Delta(1600)}},
\nonumber
\ea

\ba
& &\hspace*{-8em}\Gamma (\Delta(1600) \to \pi \Delta ) \nonumber\\
& &\hspace*{-7em} = \frac{15}{4} \frac{f_{\pi \Delta \Delta(1600)}^2}{36\pi}
\frac{(M_\Delta + M_{\Delta(1600)})^2}{m_\pi^2} \frac{M_\Delta |\vec{p}|}{M_{\Delta(1600)}} \nonumber\\
& &\hspace{-8em} \times\left[ \left(\frac{E_\Delta}{M_\Delta}\right)-1 \right]
\left[ 2\left(\frac{E_\Delta}{M_\Delta}\right)^2 - 2\left(\frac{E_\Delta}{M_\Delta}\right) +5 \right],
\label{gammaDsDpi}\\
{\rm with}\hspace{1em} |\vec{p}|
&=& \frac{\lambda^{1/2}(M_{\Delta(1600)}^2,M_\Delta^2,m_\pi^2)}{2M_{\Delta(1600)}},
\nonumber
\ea
where, $\lambda(x,y,z)\equiv x^2+y^2+z^2-2xy-2yz-2zx$.


\vspace{.15cm}
\begin{references}


\bibitem{Burkert04}
  V.~D.~Burkert and T.~S.~H.~Lee,
  Int.\ J.\ Mod.\ Phys.\  E {\bf 13}, 1035 (2004)
  [arXiv:nucl-ex/0407020].


\bibitem{Drechsel07}
  D.~Drechsel, S.~S.~Kamalov and L.~Tiator,
  Eur.\ Phys.\ J.\  A {\bf 34}, 69 (2007)
  [arXiv:0710.0306 [nucl-th]].




\bibitem{Arndt08}
  R.~Arndt, W.~Briscoe, I.~Strakovsky and R.~Workman,
  Eur.\ Phys.\ J.\  A {\bf 35}, 311 (2008).




\bibitem{Aznauryan07}
  I.~G.~Aznauryan,
  Phys.\ Rev.\  C {\bf 76}, 025212 (2007)
  [arXiv:nucl-th/0701012].



\bibitem{CLAS}
  I.~G.~Aznauryan {\it et al.}  [CLAS Collaboration],
  Phys.\ Rev.\  C {\bf 80}, 055203 (2009)
  [arXiv:0909.2349 [nucl-ex]].



\bibitem{Mokeev09}
  V.~I.~Mokeev, V.~D.~Burkert, T.~S.~H.~Lee, L.~Elouadrhiri, G.~V.~Fedotov and B.~S.~Ishkhanov,
  Phys.\ Rev.\  C {\bf 80}, 045212 (2009)
  [arXiv:0809.4158 [hep-ph]].




\bibitem{Bonn}
  A.~V.~Anisovich, E.~Klempt, V.~A.~Nikonov,
M.~A.~Matveev, A.~V.~Sarantsev and U.~Thoma,
  Eur.\ Phys.\ J.\  A {\bf 44}, 203 (2010)
  [arXiv:0911.5277 [hep-ph]].


\bibitem{Doring06}
  M.~Doring, E.~Oset and D.~Strottman,
  Phys.\ Rev.\  C {\bf 73}, 045209 (2006)
  [arXiv:nucl-th/0510015].






\bibitem{Penner02}
  G.~Penner and U.~Mosel,
  Phys.\ Rev.\  C {\bf 66}, 055211 (2002)
  [arXiv:nucl-th/0207066];
  G.~Penner and U.~Mosel,
  Phys.\ Rev.\  C {\bf 66}, 055212 (2002)
  [arXiv:nucl-th/0207069].




\bibitem{CMB}         
  T.~P.~Vrana, S.~A.~Dytman and T.~S.~H.~Lee,
  Phys.\ Rept.\  {\bf 328}, 181 (2000)
  [arXiv:nucl-th/9910012].



\bibitem{KSU}
  D.~M.~Manley,
  Int.\ J.\ Mod.\ Phys.\  A {\bf 18} (2003) 441.





\bibitem{Kamalov01}
  S.~S.~Kamalov, S.~N.~Yang, D.~Drechsel, O.~Hanstein and L.~Tiator,
  Phys.\ Rev.\ C {\bf 64}, 032201(R) (2001)
  [arXiv:nucl-th/0006068].




\bibitem{Doring09a}
  M.~Doring, C.~Hanhart, F.~Huang, S.~Krewald and U.~G.~Meissner,
  Nucl.\ Phys.\  A {\bf 829}, 170 (2009)
  [arXiv:0903.4337 [nucl-th]].



\bibitem{Julich}
  S.~Schneider, S.~Krewald and U.~G.~Meissner,
  Eur.\ Phys.\ J.\  A {\bf 28}, 107 (2006)
  [arXiv:nucl-th/0603040].


\bibitem{SatoLee}
  T.~Sato and T.~S.~H.~Lee,
  Phys.\ Rev.\  C {\bf 63}, 055201 (2001)
  [arXiv:nucl-th/0010025].


\bibitem{Diaz07c}
  B.~Julia-Diaz, T.~S.~H.~Lee, A.~Matsuyama and T.~Sato,
  Phys.\ Rev.\  C {\bf 76}, 065201 (2007)
  [arXiv:0704.1615 [nucl-th]].

\bibitem{Matsuyama07}
  A.~Matsuyama, T.~Sato and T.~S.~Lee,
  Phys.\ Rept.\  {\bf 439}, 193 (2007)
  [arXiv:nucl-th/0608051].



\bibitem{Diaz07a}
  B.~Julia-Diaz, T.~S.~H.~Lee, T.~Sato and L.~C.~Smith,
  Phys.\ Rev.\  C {\bf 75}, 015205 (2007)
  [arXiv:nucl-th/0611033].


\bibitem{Capstick92a}
  S.~Capstick,
  Phys.\ Rev.\  D {\bf 46}, 2864 (1992).






\bibitem{PDG}
  C.~Amsler {\it et al.}  [Particle Data Group],
  Phys.\ Lett.\  B {\bf 667}, 1 (2008).




\bibitem{Capstick95}
  S.~Capstick and B.~D.~Keister,
  Phys.\ Rev.\  D {\bf 51}, 3598 (1995)
  [arXiv:nucl-th/9411016].


\bibitem{Golli08b}
  B.~Golli and S.~Sirca,
  Eur.\ Phys.\ J.\  A {\bf 38}, 271 (2008)
  [arXiv:0708.3759 [hep-ph]].






\bibitem{Skorodko09b}
  T.~Skorodko {\it et al.},
  Phys.\ Lett.\  B {\bf 679}, 30 (2009)
  [arXiv:0906.3087 [nucl-ex]].


\bibitem{Skorodko10a}
  T.~Skorodko {\it et al.}  [for the CELSIUS/WASA Collaboration and for the
                  CELSIUS/WASA Collaboration a],
  arXiv:1001.5446 [nucl-ex].



\bibitem{Cao10a}
  X.~Cao, B.~S.~Zou and H.~S.~Xu,
  arXiv:1004.0140 [nucl-th].



\bibitem{Engel09}
  G.~Engel, C.~Gattringer, C.~B.~Lang, M.~Limmer, D.~Mohler and A.~Schafer,
  arXiv:0910.2802 [hep-lat].


\bibitem{Leinweber04}
  D.~B.~Leinweber, W.~Melnitchouk,
  D.~G.~Richards, A.~G.~Williams and J.~M.~Zanotti,
  Lect.\ Notes Phys.\  {\bf 663}, 71 (2005)
  [arXiv:nucl-th/0406032].



\bibitem{Erkol08}
  G.~Erkol and M.~Oka,
  Nucl.\ Phys.\  A {\bf 801}, 142 (2008)
  [arXiv:0801.0783 [nucl-th]].










\bibitem{Nucleon}
  F.~Gross, G.~Ramalho and M.~T.~Pe\~na,
  Phys.\ Rev.\  C {\bf 77}, 015202 (2008)
  [arXiv:nucl-th/0606029].



\bibitem{FixedAxis}
  F.~Gross, G.~Ramalho and M.~T.~Pe\~na,
  Phys.\ Rev.\  C {\bf 77}, 035203 (2008).


\bibitem{NDelta}
  G.~Ramalho, M.~T.~Pe\~na and F.~Gross,
  Eur.\ Phys.\ J.\  A {\bf 36}, 329 (2008)
  [arXiv:0803.3034 [hep-ph]].

\bibitem{NDeltaD}
  G.~Ramalho, M.~T.~Pe\~na and F.~Gross,
  Phys.\ Rev.\  D {\bf 78}, 114017 (2008)
  [arXiv:0810.4126 [hep-ph]].



\bibitem{LatticeD}
  G.~Ramalho and M.~T.~Pe\~na,
  Phys.\ Rev.\  D {\bf 80}, 013008 (2009)
  [arXiv:0901.4310 [hep-ph]].


\bibitem{Roper}
  G.~Ramalho and K.~Tsushima,
  Phys.\ Rev.\  D {\bf 81}, 074020 (2010)
  [arXiv:1002.3386 [hep-ph]].




\bibitem{Lattice}
  G.~Ramalho and M.~T.~Pena,
  J.\ Phys.\ G {\bf 36}, 115011 (2009)
  [arXiv:0812.0187 [hep-ph]].




\bibitem{DeltaFF0}
  G.~Ramalho and M.~T.~Pe\~na,
  J.\ Phys.\ G {\bf 36}, 085004 (2009)
  [arXiv:0807.2922 [hep-ph]]

\bibitem{DeltaFF}
  G.~Ramalho, M.~T.~Pe\~na and F.~Gross,
  Phys.\ Lett.\  B {\bf 678}, 355 (2009)
  [arXiv:0902.4212 [hep-ph]];
  G.~Ramalho, M.~T.~Pe\~na and F.~Gross,
  Phys.\ Rev.\ D {\bf 81}, 113011 (2010)
  [arXiv:1002.4170 [hep-ph]].


\bibitem{Omega}
  G.~Ramalho, K.~Tsushima and F.~Gross,
  Phys.\ Rev.\  D {\bf 80}, 033004 (2009)
  [arXiv:0907.1060 [hep-ph]].


\bibitem{Octet}
  F.~Gross, G.~Ramalho and K.~Tsushima,
  Phys.\ Lett.\  B {\bf 690}, 183 (2010)
  [arXiv:0910.2171 [hep-ph]].





\bibitem{Jones73}
  H.~F.~Jones and M.~D.~Scadron,
  Annals Phys.\  {\bf 81}, 1 (1973).




\bibitem{Becchi65}
  C.~Becchi and G.~Morpurgo,
  Phys.~Lett.\ {\bf 17}, 352 (1965).


\bibitem{Isgur82}
  N.~Isgur, G.~Karl and R.~Koniuk,
  Phys.\ Rev.\  D {\bf 25}, 2394 (1982).





\bibitem{Pascalutsa07}
  V.~Pascalutsa, M.~Vanderhaeghen and S.~N.~Yang,
  Phys.\ Rept.\  {\bf 437}, 125 (2007)
  [arXiv:hep-ph/0609004].


\bibitem{Villano09}
  A.~N.~Villano {\it et al.},
  Phys.\ Rev.\  C {\bf 80}, 035203 (2009)
  [arXiv:0906.2839 [nucl-ex]].






\bibitem{Faessler06}
  A.~Faessler, T.~Gutsche, B.~R.~Holstein, V.~E.~Lyubovitskij,
D.~Nicmorus and K.~Pumsa-ard,
  Phys.\ Rev.\  D {\bf 74}, 074010 (2006)
  [arXiv:hep-ph/0608015].



\bibitem{Rohrwild07}
  J.~Rohrwild,
  Phys.\ Rev.\  D {\bf 75}, 074025 (2007)
  [arXiv:hep-ph/0701085].


\bibitem{Braun06}
  V.~M.~Braun, A.~Lenz, G.~Peters and A.~V.~Radyushkin,
  Phys.\ Rev.\ D {\bf 73}, 034020 (2006)
  [arXiv:hep-ph/0510237].



\bibitem{Wang09}
  L.~Wang and F.~X.~Lee,
  Phys.\ Rev.\  D {\bf 80}, 034003 (2009)
  [arXiv:0905.1944 [hep-ph]].






\bibitem{Gail06}
  T.~A.~Gail and T.~R.~Hemmert,
  arXiv:nucl-th/0512082.
  Eur.\ Phys.\ J.\ A {\bf 28}, 91 (2006).

\bibitem{Pascalutsa06a}
  V.~Pascalutsa and M.~Vanderhaeghen,
  Phys.\ Rev.\ D {\bf 73}, 034003 (2006)
  [arXiv:hep-ph/0512244].






\bibitem{Dong99}
  Y.~B.~Dong, K.~Shimizu, A.~Faessler and A.~J.~Buchmann,
  Phys.\ Rev.\  C {\bf 60}, 035203 (1999).


\bibitem{Alberto04}
  P.~Alberto, L.~Amoreira, M.~Fiolhais, B.~Golli and S.~Sirca,
  Eur.\ Phys.\ J.\  A {\bf 26}, 99 (2005)
  [arXiv:hep-ph/0409246].



\bibitem{Chen08}
  D.~Y.~Chen and Y.~B.~Dong,
  Commun.\ Theor.\ Phys.\  {\bf 50}, 142 (2008).





\bibitem{ChPT}
  U.~G.~Meissner,
  AIP Conf.\ Proc.\  {\bf 904}, 142 (2007)
  [arXiv:nucl-th/0701094].







\bibitem{Alexandrou08}
  C.~Alexandrou, G.~Koutsou, H.~Neff, J.~W.~Negele, W.~Schroers and A.~Tsapalis,
  Phys.\ Rev.\  D {\bf 77}, 085012 (2008)
  [arXiv:0710.4621 [hep-lat]].





\bibitem{Savkli01}
  C.~Savkli and F.~Gross,
  Phys.\ Rev.\  C {\bf 63}, 035208 (2001)
  [arXiv:hep-ph/9911319].


\bibitem{Gross06}
  F.~Gross and P.~Agbakpe,
  Phys.\ Rev.\  C {\bf 73}, 015203 (2006)
  [arXiv:nucl-th/0411090].






\bibitem{Diaz04}
  B.~Julia-Diaz, D.~O.~Riska and F.~Coester,
  Phys.\ Rev.\  C {\bf 69}, 035212 (2004)
  [Erratum-ibid.\  C {\bf 75}, 069902 (2007)]
  [arXiv:hep-ph/0312169].







\bibitem{Aznauryan08}
  I.~G.~Aznauryan, V.~D.~Burkert and T.~S.~Lee,
  arXiv:0810.0997 [nucl-th].





\bibitem{Giannini91}
  M.~M.~Giannini,
  Rept.\ Prog.\ Phys.\  {\bf 54}, 453 (1991).




\bibitem{Arndt04}
  D.~Arndt and B.~C.~Tiburzi,
  Phys.\ Rev.\  D {\bf 69}, 014501 (2004)
  [arXiv:hep-lat/0309013].





\bibitem{Cloet03}
  I.~C.~Cloet, D.~B.~Leinweber and A.~W.~Thomas,
  Phys.\ Lett.\  B {\bf 563}, 157 (2003)
  [arXiv:hep-lat/0302008].
















\bibitem{Crawford83}
  R.~l.~Crawford and W.~t.~Morton,
  Nucl.\ Phys.\  B {\bf 211}, 1 (1983).

\bibitem{Arndt96}
  R.~A.~Arndt, I.~I.~Strakovsky and R.~L.~Workman,
  Phys.\ Rev.\  C {\bf 53}, 430 (1996)
  [arXiv:nucl-th/9509005].






\bibitem{Tsushima99}
  K.~Tsushima, A.~Sibirtsev, A.~W.~Thomas, and G.~Q.~Li,
  Phys.\ Rev.\  C {\bf 59}, 369 (1999)
  [Erratum-ibid.\  C {\bf 61}, 029903 (2000)]
  [arXiv:nucl-th/9801063].













\bibitem{Dodd81}     
  L.~R.~Dodd, A.~W.~Thomas and R.~F.~Alvarez-Estrada,
  Phys.\ Rev.\  D {\bf 24}, 1961 (1981).



\bibitem{Thomas81}    
  A.~W.~Thomas, S.~Theberge and G.~A.~Miller,
  Phys.\ Rev.\  D {\bf 24}, 216 (1981).


\bibitem{Thomas84}
  A.~W.~Thomas,
  Adv.\ Nucl.\ Phys.\  {\bf 13}, 1 (1984).


\bibitem{Thomas83}
  S.~Theberge and A.~W.~Thomas,
  Nucl.\ Phys.\  A {\bf 393}, 252 (1983).


\bibitem{Lu97}     
  D.~H.~Lu, A.~W.~Thomas and A.~G.~Williams,
  Phys.\ Rev.\  C {\bf 55}, 3108 (1997)
  [arXiv:nucl-th/9612017].


\bibitem{Lu98}
  D.~H.~Lu, A.~W.~Thomas and A.~G.~Williams,
  Phys.\ Rev.\  C {\bf 57}, 2628 (1998)
  [arXiv:nucl-th/9706019].



\bibitem{Lu01}
  D.~H.~Lu, S.~N.~Yang and A.~W.~Thomas,
  Nucl.\ Phys.\  A {\bf 684} (2001) 296.




\bibitem{Thomas07}
  A.~W.~Thomas,
  Prog.\ Theor.\ Phys.\  {\bf 168}, 614 (2007)
  [arXiv:0711.2259 [nucl-th]].





  \bibitem{CBM1}
  K.~Tsushima, T.~Yamaguchi, Y.~Kohyama and K.~Kubodera,
  Nucl.\ Phys.\  A {\bf 489}, 557 (1988).


  \bibitem{CBM2}
  T.~Yamaguchi, K.~Tsushima, Y.~Kohyama and K.~Kubodera,
  Nucl.\ Phys.\  A {\bf 500}, 429 (1989).

  \bibitem{CBM3}
  K.~Kubodera, Y.~Kohyama, K.~Oikawa and C.~W.~Kim,
  Nucl.\ Phys.\  A {\bf 439}, 695 (1985).








\bibitem{Riska01}
  D.~O.~Riska and G.~E.~Brown,
  Nucl.\ Phys.\  A {\bf 679}, 577 (2001)
  [arXiv:nucl-th/0005049].











\end{references}
\end{document}